\documentclass[10pt]{wlscirep}
\usepackage[utf8]{inputenc}
\usepackage[T1]{fontenc}
\usepackage{bm}
\usepackage{amsmath}
\usepackage{siunitx}
\usepackage{caption}
\usepackage{lineno}

\usepackage{mathrsfs}

\usepackage[acronym]{glossaries}  
\newacronym{aigc}{AIGC}{AI-Generated Content}
\newacronym{gpu}{GPU}{Graphic Processing Unit}
\newacronym{cmos}{CMOS}{Metal-Oxide-Semiconductor}
\newacronym{tia}{TIA}{Transimpedance Amplifier}
\newacronym{relu}{ReLU}{Rectified Linear Unit}
\newacronym{sde}{SDE}{Stochastic Differential Equation}
\newacronym{ode}{ODE}{Ordinary Differential Equation}
\newacronym{vae}{VAE}{Variational Autoencoder}
\newacronym{kl}{KL}{Kullback-Leibler}
\newacronym{emnist}{EMNIST}{Extended Modified National Institute of Standards and Technology}
\newacronym{eds}{EDS}{Energy-dispersive X-ray Spectroscopy}
\newacronym{tem}{TEM}{Transmission Electron Micrograph}
\glsdisablehyper

\newcommand{\figref}[1]{Fig.~{\ref{#1}}}

\title{Resistive Memory-based Neural Differential Equation Solver for Score-based Diffusion Model}

\author[1,2,5,$^{\dagger}$]{Jichang Yang}
\author[1,2,5,$^{\dagger}$]{Hegan Chen}
\author[2,$^{\dagger}$,*]{Jia Chen}

\author[1,2,5]{Songqi Wang}
\author[1,2]{Shaocong Wang}
\author[1,2]{Yifei Yu}
\author[1,2]{Xi Chen}
\author[1,2]{Bo Wang}
\author[1,2,5]{Xinyuan Zhang}
\author[1]{Binbin Cui}
\author[1,3,5,6]{Yi Li}
\author[1,2]{Ning Lin}
\author[1]{Meng Xu}

\author[7]{Yi Li}
\author[3,6]{Xiaoxin Xu}
\author[1]{Xiaojuan Qi}
\author[1,2,5,*]{Zhongrui Wang}
\author[4,*]{Xumeng Zhang}
\author[3,6,*]{Dashan Shang}

\author[1,5]{Han Wang}
\author[4]{Qi Liu}
\author[2,8]{Kwang-Ting Cheng}
\author[3,4]{Ming Liu}

\affil[1]{Department of Electrical and Electronic Engineering, the University of Hong Kong, Hong Kong, China}
\affil[2]{ACCESS – AI Chip Center for Emerging Smart Systems, InnoHK Centers, Hong Kong Science Park, Hong Kong, China}
\affil[3]{Laboratory of Microelectronic Devices and Integrated Technology, Institute of Microelectronics, Chinese Academy of Sciences, Beijing 100029, China}
\affil[4]{State Key Laboratory of Integrated Chips and Systems, Frontier Institute of Chip and System, Fudan University, Shanghai 200433, China}
\affil[5]{Institute of the Mind, the University of Hong Kong, Hong Kong, China}
\affil[6]{University of Chinese Academy of Sciences, Beijing 100049, China}
\affil[7]{School of Integrated Circuits, Hubei Key Laboratory for Advanced Memories, Huazhong University of Science and Technology, Wuhan 430074, China.}
\affil[8]{Department of Electronic and Computer Engineering, the Hong Kong University of Science and Technology, Hong Kong, China}

\affil[$^{\dagger}$]{These authors contributed equally.}
\affil[*]{e-mail: zrwang@eee.hku.hk; jjiachen@ust.hk; xumengzhang@fudan.edu.cn;  shangdashan@ime.ac.cn}

\begin{abstract}

Human brains image complicated scenes when reading a novel. Replicating this imagination is one of the ultimate goals of \gls{aigc}. However, current \gls{aigc} methods, such as score-based diffusion, are still deficient in terms of rapidity and efficiency. 
This deficiency is rooted in the difference between the brain and digital computers. Digital computers have physically separated storage and processing units, resulting in frequent data transfers during iterative calculations, incurring large time and energy overheads. This issue is further intensified by the conversion of inherently continuous and analog generation dynamics,  which can be formulated by neural differential equations, into discrete and digital operations. 
Inspired by the brain, we propose a time-continuous and analog in-memory neural differential equation solver for score-based diffusion, employing emerging resistive memory. The integration of storage and computation within resistive memory synapses surmount the von Neumann bottleneck, benefiting the generative speed and energy efficiency. The closed-loop feedback integrator is time-continuous, analog, and compact, physically implementing an infinite-depth neural network. Moreover, the software-hardware co-design is intrinsically robust to analog noise.
We experimentally validate our solution with 180~nm resistive memory in-memory computing macros. Demonstrating equivalent generative quality to the software baseline, our system achieved remarkable enhancements in generative speed for both unconditional and conditional generation tasks, by factors of 64.8 and 156.5, respectively. Moreover, it accomplished reductions in energy consumption by factors of 5.2 and 4.1. Our approach heralds a new horizon for hardware solutions in edge computing for generative AI applications.

\end{abstract}

\begin{document}

\flushbottom
\maketitle


\section*{Introduction}

Imagination is an intricate cognitive process that showcases human brain's capacity for synthesizing memory, experiential knowledge, and creative thought\cite{mullally2014memory}, with remarkable efficiency and speed. Emulating this ability is a central goal within the domain of \gls{aigc}, with broad and promising applications spanning video game design, scenario visualization, and metaverse infrastructure. The recent development of the OpenAI's Sora\cite{videoworldsimulators2024} marks a significant advancement, offering a text-to-video generation facility that approximates the vividness of human imagination in producing realistic videos. The core algorithm under Sora's functionality is the diffusion model\cite{ho2020denoising,karras}, an algorithm that formulates a neural differential equation for the transformation of one probability distribution into another\cite{SD, ViD4k,a1,a3,a9,RN94,a2,a6_3d}.

Despite their advances, diffusion models still lag in power consumption and generation speed on standard computers compared to the human brain, the latter comes up complicated imagination (e.g. when reading a novel) instantly while consumes about \SI{20}{\watt}. The essence of score-based diffusion models lies in a neural differential equation\cite{song2020score}. The process of video or image synthesis begins with an initial sampling from a noise distribution, followed by iterative resolution of the neural differential equation to construct the final video or image. This encounters two challenges on digital computers: 
(1) The sampling process of diffusion models involves iteratively solving neural differential equations, or a substantial number of network inferences\cite{b1,b2,b4,b3,b6,b11,b12,b13}. This yields considerable time and energy overheads on digital computers due to the massive data shuttling between physically separated memory and processing units in the von Neumann architecture\cite{rc2}, along with the \gls{cmos} device sizes that are nearing their physical limits\cite{s1,s5,s9,s10, sun2023full}. 
(2) Digital computers use numerical methods to discretize and digitize time-continuous analog signals, which introduces truncation and round-off errors that can only be mitigated by further increasing the discrete steps, or the count of network inferences, such measures result in additional time and energy expenditures\cite{acc2,acc3,acc4,acc5,acc6,acc10,acc11}.

To address this, we have developed a time-continuous and analog in-memory neural differential equation solver using resistive memory for score-based diffusion model, featuring the following advantages:
(1) In-memory computing with resistive memory mitigates the von Neumann bottleneck\cite{rc1,rc4,rc5,rc6,rc9,yan2023memristive}. Resistive memory arrays physically implement analog synaptic weight matrices of neural networks. Each memory cell not only stores the synaptic weight using analog conductance, but also computes right at where the data is store using Ohm’s law for multiplication and Kirchhoff’s current law for summation. This brain-inspired in-memory computing significantly obviate the energy and time overheads in von Neumann computers\cite{rc7,rc8,SVM,rc10,s6,s7,s8,s11,s12,moon2019temporal,yang2020retransformer}. In addition, resistive memory cells have a simple, capacitor-like structure, equipping them with excellent scalability and three-dimensional (3D) stackability\cite{zhu2020comprehensive}.
(2) The resistive memory-based analog neural network, incorporating the closed-loop feedback integrator, presents a time-continuous, analog solution to the score-based diffusion model, effectively implements an infinite depth neural network. This method obviates the need for discretization and digitization, thereby eliminating the truncation and round-off errors that are intrinsic to numerical methods in digital platforms.

In this article, we validated our approach with a \SI{180}{\nano\meter} resistive memory in-memory computing macro, for both unconditional generation and classifier-free guidance\cite{CFG} latent diffusion generation. Compared to state-of-the-art digital hardware, under the same generation quality, our system manifests a 64.8x and 156.5x increase in sampling speed for the unconditional and conditional latent diffusion, respectively. Furthermore, the system achieves a power consumption reduction of 80.8\% and 75.6\% for each task, respectively. 
Moreover, the score-based diffusion model incorporates stochasticity for diversity, which partially leverages the analog circuit noise\cite{noise2,yang2022tolerating,wei2023three,yan2023memristive,cai2020power}, such as the resistive memory temporal conductance fluctuation. Our system resonates with the time-continuous, analog, and stochastic characteristics of neural processes in human imagination, which paves the way for future brain-like \gls{aigc} systems at the edge.

\section*{Brain-inspired software-hardware co-design}

\begin{figure}[!t]
    \centering
    \includegraphics[width=1 \linewidth]{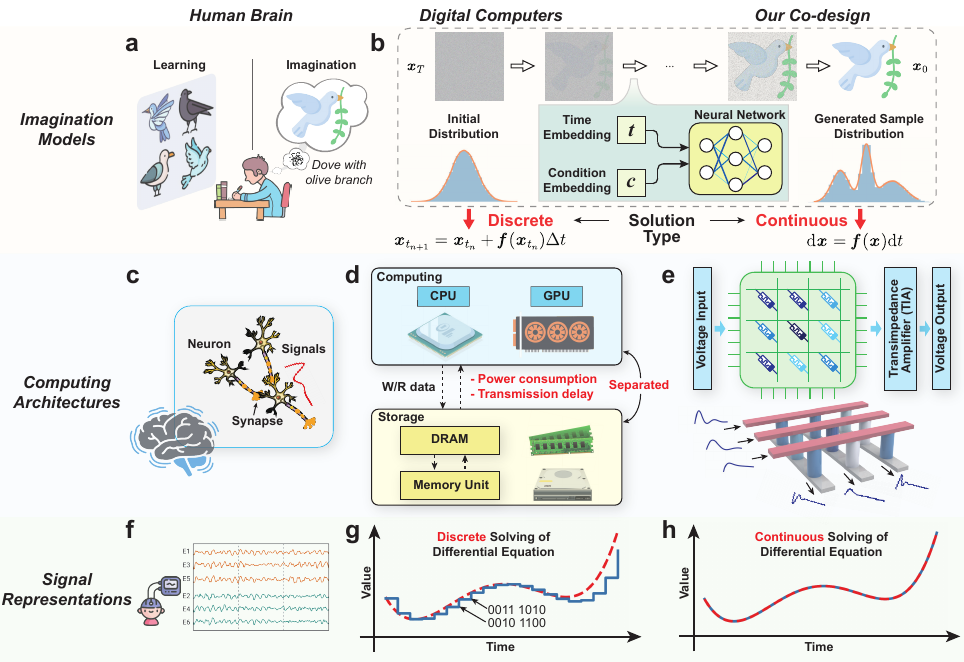}
    \caption{
        \textbf{Comparison of imagination model, computing architecture, and signal representation between the human brain, digital computer and our system.} 
        \textbf{a-c,} Imagination models.
        \textbf{a,} Human imagination. Human beings possess the innate ability to imagine (e.g. blue dove with olive branch) after learning relevant knowledge and experiences.
        \textbf{b,} Generative diffusion model. Both digital computer and our system samples from a standard Gaussian distribution and transforming it into samples from the target distribution through progressive transformation by a neural differential equation.
        \textbf{c-e,} Comparison of computing architectures.
        \textbf{c,} Biological neural network of the brain. Synapses store synaptic strength and modulate signal transmission between adjacent neurons according to the strength.
        \textbf{d,} Von Neumann architecture. Conventional digital computers feature physically separated storage and computing units, which suffers from high power consumption and low speed due to data shuttling.
        \textbf{e,} Resistive memory-based in-memory computing. Inspired by the brain, resistive memory cells emulate synaptic functionality by concurrently storing and processing information in an energy-efficient manner.
        \textbf{f-g,} Comparison of signal representations.
        \textbf{f,} Human brainwaves. The signals in the brain are time-continuous and analog.
        \textbf{g,} Signals of digital computers. Discretization and digitization introduce truncation and round-off errors.
        \textbf{h,} Signals of our system. Like the brain, our system operates with fully time-continuous and analog signals.
    }
    \label{fig1}
\end{figure}

\figref{fig1} compares human brain, digital computers, and our co-designed system across three different aspects: imagination models, computing architectures, and signal representations.

Regarding the paradigm of imagination, \figref{fig1}a depicts the innate ability of humans to imagine. Upon comprehending certain attributes of an object through sensory experiences such as observation and auditory perception, individuals can recall or conceptualize something entirely novel\cite{modell2003imagination}. For instance, when reading a novel, readers are capable of envisioning the scenes within their mental landscape.
Generation models, such as diffusion model, aim to mirror the imagination of humans. \figref{fig1}b schematically illustrates the image generation according to textual descriptions.
At their core, diffusion models function by learning a denoising model that progressively removes noise from random samples drawn from a standard Gaussian distribution to generate the target images\cite{ho2020denoising}.
The process is modulated by the incorporation of temporal and conditional information into the neural network during generation.

In terms of computing architecture, \figref{fig1}c depicts the human brain, portrays the human brain as a sophisticated biological neural network, comprised of neurons interconnected by synapses. Synapses both store information in terms of synaptic strength (i.e. neurotransmitter levels and receptor sensitivity) and modulate the signal transmission between neurons. A neuron integrates incoming spiking signals and produces its own spike once reaching a threshold. Benefiting from this in-memory computing architecture, the human brain is capable of imagining in a fast and low-power manner.
However, as shown in \figref{fig1}d, existing digital computers are primarily based on the von Neumann architecture where memory and computing units are separated physically. This separation often incurs significant temporal and energy costs due to the extensive data transfer required between these components.
\figref{fig1}e illustrates our resistive memory-based in-memory computing\cite{rr2,s2,s3,s4,wang2023analog,ielmini2021brain}, which resembles the human brain. The resistive memory cells work as synapses which store synaptic strength in terms of conductance and modulate signal transmission using Ohm's law, while neural summation is carried out by Kirchhoff's law. The collocation of memory and processing considerably benefits in computation speed and energy efficiency\cite{rr1,c12,c1,rr2,yi2023activity,sun2020one,zhang2023edge,milo2021accurate}.

In terms of signal representation, as demonstrated in \figref{fig1}f, brainwaves are inherently time-continuous and analog during imagination and cognition.
In contrast, traditional digital computers are driven by clock signals to execute Boolean operations via \gls{cmos} logic gates, therefore temporal signals are discretized and digitized as shown in \figref{fig1}g. This inevitably introduces truncation and round-off errors when approximating time-continuous and analog signals.
On the other hand, our resistive memory-based neural differential equation solver follows the signal representation in human brain. As depicted in \figref{fig1}h, our hardware is fully time-continuous and analog, which physically implements and accelerates score-based diffusion model, thereby circumventing the errors commonly introduced by discrete numerical computations.

\section*{Resistive memory-based neural differential equation solver}

\begin{figure}[!t]
    \centering
    \includegraphics[width=1 \linewidth]{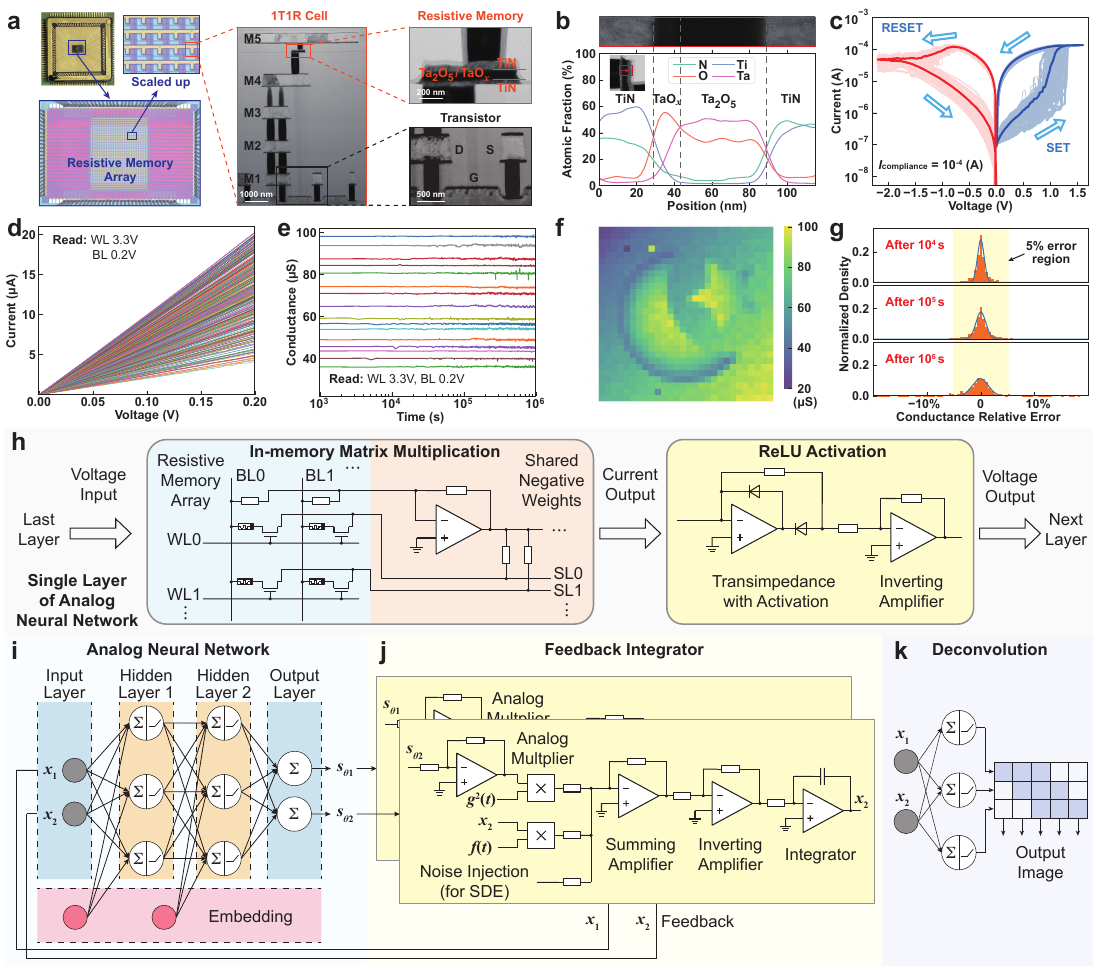}
    \caption{
        \textbf{Resistive memory characterization and circuit design of neural differential equation solver.} 
        \textbf{a-g,} Physical and electrical characterization of resistive memory.
        \textbf{a,} Optical photos of the resistive memory array, cross-sectional \gls{tem} of a single 1-transistor-1-resistor cell (scale bar: \SI{1000}{\nano\meter}), $\mathrm{TiN}/\mathrm{TaO}_x/\mathrm{Ta}_2\mathrm{O}_5/\mathrm{TiN}$ resistive memory cell (scale bar: \SI{200}{\nano\meter}) and the series \SI{180}{\nano\meter} transistor (scale bar: \SI{500}{\nano\meter}). 
        \textbf{b,} \gls{eds} line profile of a single resistive memory cell.
        \textbf{c,} Quasi static I-V sweeps of the a resistive memory cell, showing repeatable bipolar resistive switching behavior.
        \textbf{d,} More than 64 discernible linear conductance states of a resistive memory cell.
        \textbf{e,} More than $10^{6}$s retention of different conductance states of resistive memory cells.
        \textbf{f,} 32$\times$32 resistive memory array programmed to display a moon and star conductance pattern.
        \textbf{g,} Resistive memory arrays conductance error distribution at different times.
        \textbf{h-k,} Circuit blocks of the neural differential equation solver.
        \textbf{h,} Diagram of a single-layer analog neural network, where resistive memory cells form differential pairs with row-shared negative weights circuit. Input voltages are applied to the BLs of a resistive memory array, with resultant output currents traversing through SLs. The \gls{relu} activation is physically implemented by \gls{tia} and inverting amplifiers.
        \textbf{i,} The analog neural network cascades three single layer analog neural network modules.
        \textbf{j,} The output of the analog neural network is received by the feedback integrator circuit, consisting of an analog integrator, which in turn provides feedback to the input of the analog neural network, forming a closed-loop circuit to solve neural differential equations. 
        \textbf{k,} In latent diffusion, the solution provided by the neural differential equation solver is translated into pixel space by resistive memory-based deconvolution decoder.
    }
    \label{fig2}
\end{figure}

\figref{fig2} shows the design of individual resistive memory devices and circuit blocks of our system.

Device-wise, \figref{fig2}a,b show the physical characterization of resistive memory. \figref{fig2}a presents from left to right: optical images of the resistive memory array, a Transmission Electron Micrograph (\gls{tem}) image of a 1-transistor-1-resistor cell, \gls{tem} images of a $\mathrm{TiN}/\mathrm{TaO}_x/\mathrm{Ta}_2\mathrm{O}_5/\mathrm{TiN}$ resistive memory cell and the transistor in series using \SI{180}{\nano\meter} technology node. The memory cells arranged in each row are interconnected by a common Word Line (WL) and Source Line (SL), which link the gate and source terminals of the transistors, respectively. Conversely, memory cells arranged in columns share a common Bit Line (BL). \figref{fig2}b validates the chemical composition of the resistive memory cell using the \gls{eds} line profile.

\figref{fig2}c-g present the electrical characterization of resistive memory. The 200-cycle quasi-static I-V sweeps (\figref{fig2}c) show highly uniform bipolar resistive switching, where the solid lines correspond to the average current. 
\figref{fig2}d presents more than 64 distinguishable linear conductance states of the device within the conductance range between \SI{0.02}{\milli\siemens} and \SI{0.10}{\milli\siemens}, showcasing the multilevel data storage capability of the resistive memory cell. 
\figref{fig2}e demonstrates that the resistive memory cells maintain relatively stable programmed conductance states over time, with small temporal fluctuations or read noise. 
\figref{fig2}f presents the programmed moon and star analog conductance pattern on a resistive memory array, which shows the good yield and reasonable programming accuracy at the array level (programming logic diagram shown in Supplementary Fig.~3). 
\figref{fig2}g shows the conductance error distribution within the resistive memory array. The conductance relative errors follow a Gaussian distribution and do not exhibit significant temporal variation.

Circuit-wise, \figref{fig2}h-k illustrate the circuit of each block in our system. In the context of a diffusion model, its generative process involves solving the following \gls{sde}\cite{song2020score}:

\begin{equation}
    \frac{\mathrm{d}\boldsymbol{x}}{\mathrm{d} t} = \boldsymbol{f}(\boldsymbol{x}, t) - g^2(t) \boldsymbol{s}_\theta(\boldsymbol{x},t) + g(t) \frac{\mathrm{d} \boldsymbol{w}}{\mathrm{d} t} \
     = \boldsymbol{F}_{\mathrm{SDE}}(\boldsymbol{x}, t)
\label{SDE}
\end{equation}

When utilizing the \textit{probability flow} method for sampling, it is not necessary to inject noise, and the generative process is reduced to solving the following \gls{ode}:

\begin{equation} 
    \frac{\mathrm{d}\boldsymbol{x}}{\mathrm{d} t} = \boldsymbol{f}(\boldsymbol{x}, t) - \frac{1}{2} g^2(t) \boldsymbol{s}_\theta(\boldsymbol{x},t) = \boldsymbol{F}_{\mathrm{ODE}}(\boldsymbol{x}, t)
\label{ODE}
\end{equation}

$\boldsymbol{F}_{\mathrm{SDE}}(\boldsymbol{x}, t)$ and $\boldsymbol{F}_{\mathrm{ODE}}(\boldsymbol{x}, t)$ are the differential terms for \gls{sde} and \gls{ode} respectively. The solutions are obtained by calculating the following integral form:

\begin{equation} 
    \boldsymbol{x}_0 = \int_{T}^{0} \boldsymbol{F}_{\mathrm{SDE/ODE}}(\boldsymbol{x}, t) \mathrm{d} t , \
    \quad \mathrm{with} \quad \boldsymbol{x}_T \sim N(\mathbf{0}, \boldsymbol{I}_n)
\label{int}
\end{equation}

Here the term $\boldsymbol{s}_\theta(\boldsymbol{x},t)$ is the score function to be parameterized by the analog  neural network \figref{fig2}h-i, while $\boldsymbol{f}(\boldsymbol{x}, t)$ and $g(t)$ serve as the model's forward process drift and diffusion components, respectively. $w$ denotes the standard random process (Wiener stochastic process). $n$ represents the dimension of $\boldsymbol{x}$. In this setup, $\boldsymbol{f}(\boldsymbol{x}, t)$ is defined as a linear function of the form $f(t)\boldsymbol{x}$. Both $f(t)$ and $g^2(t)$ are crafted as predetermined analog signals (definition shown in Method).

\figref{fig2}h depicts the circuit schematic of a single-layer analog neural network, including an in-memory matrix multiplication module and a nonlinear activation function module. The matrix multiplication module is composed of a resistive memory array and a summing amplifier, serving as the shared negative weights, so each resistive memory cell is equivalently paired with a fixed differential resistor of \SI{20}{\kilo\ohm}, which saves 50\% of silicon area compared to ordinary differential pairs. The actual conductance weight of each differential pair is $G=G_{\mathrm{mem}}-G_{\mathrm{fixed}}$, with a range of approximately \SI{-0.03}{\milli\siemens} to \SI{0.05}{\milli\siemens}. The input voltage vector is multiplied by the conductance matrix in the resistive memory, and the resulting output current is processed by Transimpedance Amplifiers (\gls{tia}s) and inverting amplifiers to generate the output voltage vectors. 
The nonlinear activation module consists of two diodes paired with a \gls{tia} to clamp the output voltage upper limit to \SI{0}{\volt}, physically implementing the rectified linear units (\gls{relu})\cite{fully}.

\figref{fig2}i shows the diagram of a multi-layer analog neural network that cascades several single-layer analog neural network modules. The time and conditional embedding for the intermediate layers of the neural network can be accomplished by injecting specific amounts of current into \gls{tia}s. 

\figref{fig2}j illustrates the diagram of a feedback integrator circuit. Analog multipliers and summing amplifiers first perform multiplications and additions, respectively, physically implementing the operations of $\boldsymbol{F}_{\mathrm{SDE}}(\boldsymbol{x}, t)$ and $\boldsymbol{F}_{\mathrm{ODE}}(\boldsymbol{x}, t)$. Then analog integrator, composed of operational amplifier and capacitor, is used to obtain the integrated voltages. 
This integrated output vector is fed back into the analog neural network input, forming a closed-loop system that is equivalent to Eq.~(\ref{int}). By pre-charging the capacitors in the integrator circuit, the initial conditions of the system are set, allowing for the commencement of the autonomous resolution of the neural differential equations.
For applications involving latent diffusion, it's necessary to map generated samples from the latent space encoded by a \gls{vae} back into the pixel space. This is accomplished using the a deconvolution circuit shown in \figref{fig2}k, which also employs resistive memory arrays for matrix-vector multiplications.

\section*{Unconditional generation of circular distribution}

\begin{figure}[!t]
    \centering
    \includegraphics[width=1 \linewidth]{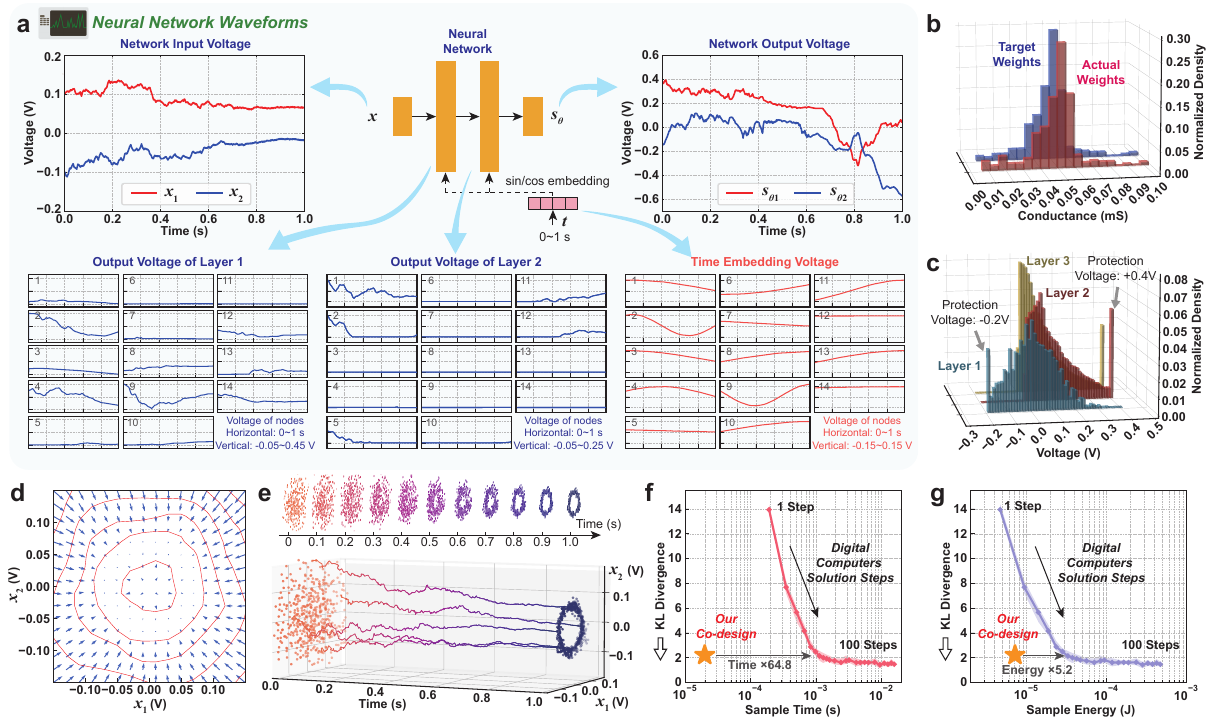}
    \caption{
        \textbf{Experimental demonstration of unconditional circular distribution generation.} 
        \textbf{a,} Experimental voltage waveforms of the analog neural network during a single sampling.
        \textbf{b,} Histogram of the offline optimized analog neural network weights and the experimental weights programmed into the resistive memory arrays.
        \textbf{c,} Histogram of the input voltages for each layer of the analog neural network.
        \textbf{d,} Schematic illustration of the two-dimensional gradient vector field output of the analog neural network. The axes represent the input voltages to the neural network, while the arrows depict the output voltage vectors from the analog neural network, with the isopotential curves denoted in red.
        \textbf{e,} Time slices of the two-dimensional distribution of analog neural network input vector from 1000 times sampling, along with waveforms showing the example trajectories over time of sampling. Initial voltage vectors, drawn from a two-dimensional Gaussian distribution, evolve to achieve the intended circular distribution after a predefined resolution period.
        \textbf{f,} Comparison of the generation speed of our co-design with that of a state-of-the-art digital hardware, showing 64.8x improvement under the same generation quality.
        \textbf{g,} Comparison of the energy consumption of the analog and a state-of-the-art digital hardware, showing a 80.8\% decrement.
    }
    \label{fig3}
\end{figure}

We first validated our fully hardware design on an unconditional generation task, to generate a circular distribution within a two-dimensional plane (system details shown in Supplementary Fig.~5). The data is two-dimensional, and a three-layer fully connected analog neural network was employed to parameterize the score function in Eq.~(\ref{SDE}) or Eq.~(\ref{ODE}). The noise injected into each dimension was $g(t)\epsilon$, with $\epsilon \sim N(0, 1)$, to approximate desired noise. Notably, in the score-based diffusion algorithm, the reverse generation process begins at time $T$ and concludes at time $0$\cite{song2020score}. Within the experimental setup, the recorded time $t = \SI{0}{\second}$ is equivalent to the initial time $T$ in the algorithm's context. Sampling commenced at $t = \SI{0}{\second}$, and stop by $t = \SI{1}{\second}$. An analog voltage of \SI{0.1}{\volt} was designated as the base unit of 1 within the software framework.

Voltage signals from various parts of the analog neural network in the sampling process are recorded for demonstration (\figref{fig3}a). The initial voltage vector (\SI{0.1}{\volt}, \SI{-0.1}{\volt}) is sampled from two-dimensional Gaussian distribution. The top-left (top-right) subplot presents the experimental voltage waveform of the input (output) of the analog neural network. The three subplots below correspond to the experimental voltage waveforms of individual neurons in the two hidden layers and the time embedding signals, respectively. \gls{relu} activation circuits restrict the outputs of some neurons to zero. The time-embedding signals are obtained by sinusoidal encoding of the time. It is observed that the input voltage vector $\boldsymbol{x}_t$ randomly sampled from a Gaussian distribution is transformed into a vector falling into the target circular distribution by $t = \SI{1}{\second}$, thereby completing a generation. 

The weights of the analog neural network are optimized offline before being deployed on resistive memory. The histogram in \figref{fig3}b compares the target conductance with the actual conductance, showing high programming precision. To prevent the applied voltages from exceeding resistive memory programming threshold, voltages are capped to the range from \SI{-0.2}{\volt} to \SI{0.4}{\volt}. \figref{fig3}c shows the histogram of the voltages applied to each layer's resistive memory cells when the neural network's initial voltage input follows a normal distribution, illustrating the effect of voltage clamping. Details of the protective circuit design are provided in Supplementary Fig.~2. \figref{fig3}d presents the experimentally measured vector field of the analog neural network's outputs in a two-dimensional plane at $t = \SI{0.5}{\second}$. The output of the analog neural network acts as a gradient field according to Eq.(\ref{SDE}) or Eq.(\ref{ODE}), guiding the voltage vector $\boldsymbol{x}_t$ towards convergence on the target circular distribution. \figref{fig3}e illustrates the standard Gaussian distribution of analog neural network's starting points, from 1000 times experimental sampling, which progressively converges onto the target circular distribution as revealed by both the time slices and trajectories.

\figref{fig3}f,g showcase the comparison between the sampling speed and energy consumption of our system and those of state-of-the-art \gls{gpu}. We use \gls{kl} divergence to measure the similarity between the generated distribution and the ground truth distribution (see Method for definition). The smaller its value, the higher the similarity and the better the quality of generation. For a projected fully integrated system, under the same generation quality, our co-design has a solution time \SI{20}{\micro\second} for a single point sampling, which represents an increase in sampling speed by a factor of 64.8 compared to the \gls{gpu}. In addition, our co-design consumes \SI{7.2}{\micro\joule} of energy per sample point, which is \textasciitilde80.8\% lower than the energy consumption of digital computers when scaled to the same technology node\cite{comp}.

\section*{Conditional generation of letters in latent space}

\begin{figure}[!t]
    \centering
    \includegraphics[width=1 \linewidth]{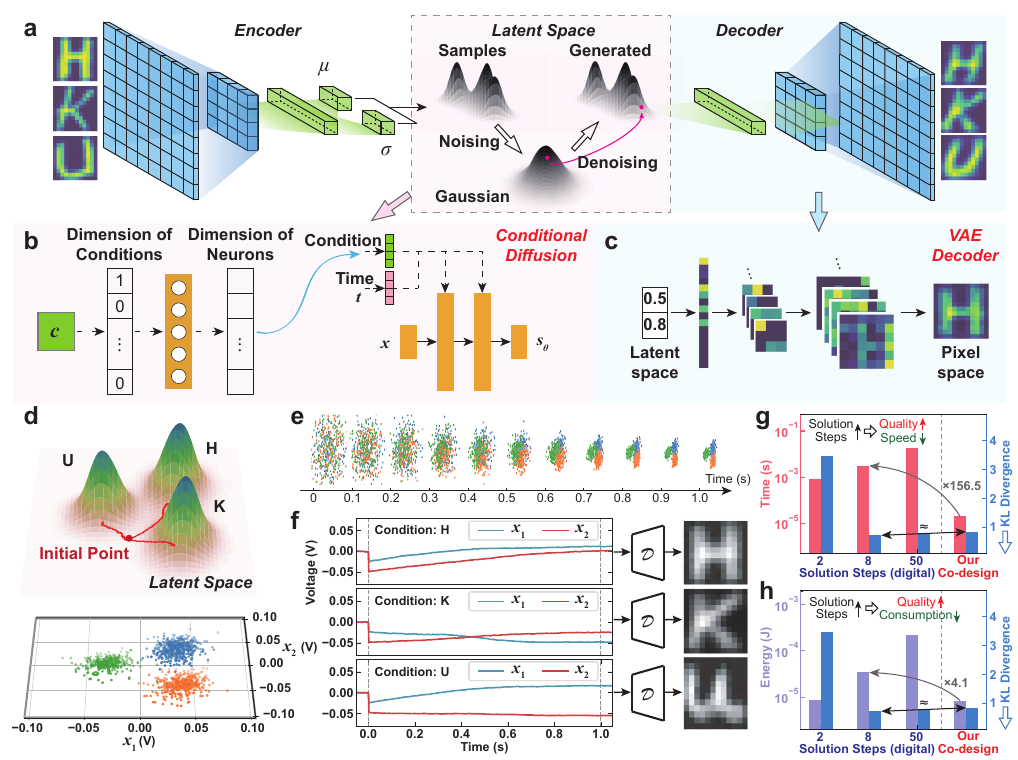}
    \caption{
        \textbf{Experimental demonstration of conditional generation of handwritten letters using latent diffusion.} 
        \textbf{a,} Software framework. An outer \gls{vae}, trained on the \gls{emnist} dataset, encodes images into a two-dimensional latent space. After generation within the latent space via conditional score-based diffusion, the \gls{vae} model's decoder transforms the latent vectors back into images.
        \textbf{b,} Condition embedding. Category labels are one-hot encoded and then transformed using random projection. The output, matching the intermediate layer dimension of the neural network, is summed with a time-encoded signal and fed into the analog neural network to guide the generation.
        \textbf{c,} Example feature maps of the \gls{vae} decoder in mapping latent vectors back to pixel image.
        \textbf{d,} Experimental distribution of three categories of handwritten letters within the latent space, each category consists of 500 times sampling.
        \textbf{e,} Time evolution of the three conditional distributions in the two-dimensional latent space.
        \textbf{f,} Experimental voltage waveforms showing different diffusion trajectories from same initial voltage latent vector under different conditions, which are subsequently decoded into handwritten letters of the corresponding categories.
        \textbf{g,} Comparison of the generation speed between our co-design and a state-of-the-art digital hardware.
        \textbf{h,} Comparison of the energy consumption between our co-design and state-of-the-art digital hardware.
    }
    \label{fig4}
\end{figure}

We then applied our co-design to the task of conditional generation for selected letters from the \gls{emnist} dataset using latent diffusion. \figref{fig4}a presents the software framework. We commenced by encoding images of the handwritten letters \textit{H}, \textit{K} and \textit{U} into a two-dimensional latent space using a pre-trained \gls{vae}. Subsequently, we employed a conditional score-based diffusion model to control the output distributions, thereby generate a specific letter. After denoising diffusion, the final latent vectors are then decoded into images using the \gls{vae} decoder. \figref{fig4}b depicts the condition embedding for the a three-layer analog neural network which parameterizes the score-based function. Here we adopted a classifier-free guidance approach\cite{CFG}, wherein the control condition embedding vector is summed with the time embedding vector, before being physically received by the three-layer analog neural network as neuron bias. The \gls{vae} decoder consists of one linear layer and two deconvolution layers, with \figref{fig4}c showing example feature maps in mapping a latent vector to handwritten letter \textit{H}.

During the training of the \gls{vae}\cite{VAE}, the distribution of images with three different labels in the latent space is controlled by pre-defined distribution centers in the latent space for each category (see Method for loss function definition). \figref{fig4}d-f show the experimental results of conditional diffusion task, \figref{fig4}d illustrates the generated three distributions under three conditions, where each condition consists of 500 times of sampling labelled in a particular color. The probability density functions of the three generated distributions are represented in the 3D plot. \figref{fig4}e shows the time evolution of the three conditional distributions in the two-dimensional latent space. Starting from an initial Gaussian distribution, it transforms into three separate distributions under different conditions. This is corroborated by \figref{fig4}f showing the trajectories in \gls{ode} score-based diffusion. Given the same initial latent coordinate (\SI{-0.025}{\volt}, \SI{-0.050}{\volt}) at $t = \SI{0}{\second}$, the different diffusion conditions produce different trajectories towards respective categorical distributions at $t = \SI{1}{\second}$, which are subsequently decoded to images of \textit{H}, \textit{K} and \textit{U}.

\figref{fig4}g,h present comparisons of the computation speed, generation quality, and power consumption between our co-design and a state-of-the-art digital hardware. For the numerical methods employed in the software baseline, using a higher number of discrete steps can lead to improved generation quality. However, this also results in increased sampling time and energy consumption. Under the same generation quality, the speed for a single sampling has been increased by 156.5$\times$ on our projected fully integrated system. Correspondingly, the energy consumption for the sampling has been reduced by 75.6\%.

\section*{Robustness of analog noise}

\begin{figure}[!t]
    \centering
    \includegraphics[width=1 \linewidth]{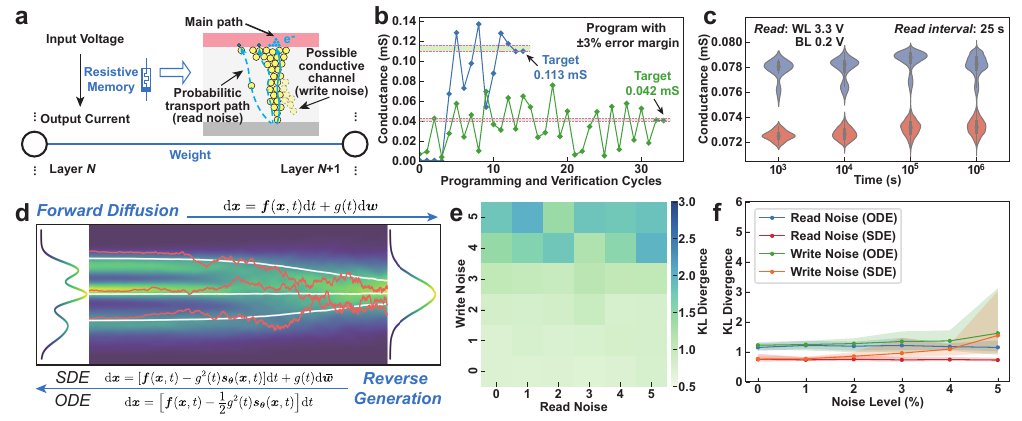}
    \caption{
        \textbf{Resistive memory noises and their impact on generation performance.} 
        \textbf{a,} Physical mechanisms of write and read noise in resistive memory.
        \textbf{b,} Experimental write noise in resistive memory programming.
        \textbf{c,} Experimental read noise over different time scales of resistive memory.
        \textbf{d,} Illustration of noise injection in score-based diffusion using \gls{sde}.
        \textbf{e,} Impact of various degrees of read noise and write noise on the generation quality.
        \textbf{f,} Respective impacts of various levels of write and read noise on the generation quality using \gls{ode} and \gls{sde} score-based diffusion.
    }
    \label{fig5}
\end{figure}


Analog computing features inevitable noise\cite{noise2, noise1}. In our system, the major noises are resistive memory write and read noises, as shown in \figref{fig5}a. Write noise stems from the inherent unpredictability of atomic movement during the redox processes that drive resistive memory switching, leading to variability in the creation of conductive filaments. Read noise, on the other hand, is attributed to conductance variability in resistive memory, which arises from thermal fluctuations and random telegraph noise affecting charge transport. Both noises impact on the precision of weights in analog neural networks.

\figref{fig5}b shows the experimentally observed write noise during iterative programming, with the green area indicating the target range. During the programming of a resistive memory to target conductance, the resistive memory cell undergoes repeated SET-RESET cycles until the conductance falls within a targeted range close to the desired value. Both the number of programming cycles required and the final conductance achieved within the target range exhibit a degree of randomness. \figref{fig5}c depicts the observed read noise. Violin chart of each color illustrates the statistical distribution of the conductance of a resistive memory measured over different time periods. It is observed that the conductance of the resistive memory fluctuates continuously. Also, the magnitude of read noise varies with mean conductance.

Score-based models incorporate noise by integrating a Wiener stochastic process\cite{song2020score} into its reverse \gls{sde} (trajectories shown in \figref{fig5}d). Due to the innate use of noise for better generation diversity, they naturally exhibit robustness to analog circuit noise. \figref{fig5}e,f shows the simulated impact of different magnitudes of noise on the similarity between the distribution generated by the diffusion model and the ground truth distribution, measured by the \gls{kl} divergence. The magnitude of noise is represented by the standard deviation of the noise added to the original weights. It is apparent that within a certain range, the read and write noise do not significantly affect the diffusion model's generation quality. This robustness is especially pronounced in the case of read noise, which is equivalent to the Wiener stochastic term of the \gls{sde} in score-based diffusion.

\section*{Discussion}

In this work, we present a time-continuous and analog neural differential equation solver that leverages resistive memory-based in-memory computing, which physically implements both unconditional and conditional latent diffusion models. 
By utilizing resistive memory for in-memory computing, our approach effectively circumvents the von Neumann bottleneck, addressing the memory wall and power wall limitations that are endemic to traditional digital computing architectures.
The time-continuous and analog differential equation solver gets rid of the errors associated with discretization and digitization in digital computers. Moreover, the diffusion model exhibits robustness against resistive memory weight noise. 
This characteristic of our system significantly improves the sampling speed of diffusion model and markedly reduces the energy consumption required for generation tasks. Our work may lay the groundwork for future fast and efficient generative edge AI.

\section*{Method}

\subsection*{Fabrication of random resistive memory chips}

In our study, the resistive memory chip was engineered using the \SI{180}{\nano\meter} technology node, featuring a 32×32 1T1R macro. The stacking-layer memory cells were fabricated on the drain terminal of the transistors and embedded between the Metal4 and Metal5 layers during the backend-of-line process. Each cell consisted of TiN electrodes as the Bottom Electrode (BE) and Top Electrode (TE), with a dual TaO$_x/$Ta$_2$O$_5$ dielectric layer sandwiched in between. Specifically, a hole with a diameter of \SI{700}{\nano\meter} was patterned by photolithography and etching on the Via4 layer, subsequently filled with TiN (\SI{40}{\nano\meter}) by Physical Vapor Deposition (PVD). The dual TaO$_x/$Ta$_2$O$_5$ dielectric layer (\SI{60}{\nano\meter}) was then deposited by Atomic Layer Deposition (ALD). Finally, the TiN-based TE (\SI{40}{\nano\meter}) was formed, followed by the top Metal5 layer deposition. The 1T1R cells in the same row shared the Word Line (WL) and Source Line (SL), connecting the gate and source terminals of the transistors, respectively. Meanwhile, cells in the same column shared the Bit Line (BL), connecting the TE terminals of the resistive memory cells, forming a crossbar structure. The device was subjected to a post-annealing treatment at \SI{400}{\celsius} for 30 minutes under vacuum conditions, which enhanced the chip's performance, yielding devices with high endurance and reliability.

\subsection*{The time-continuous and analog in-memory neural differential equation solver}

The in-memory neural differential equation solver is composed of two primary sections: an analog neural network block with resistive memory arrays and the subsequent analog feedback integrator module. On the integrated printed circuit board (PCB), the whole system is controlled by an ARM MCU STM32F407ZGT6 from STMicroelectronics. The operational amplifiers (OPAs) are OPAx171 from Texas Instruments (TI). Specifically, the OPA4171 is utilized in circuits where multiple OPAs are required to operate in parallel. For circuits with fewer OPA requirements, the OPA2171 is implemented. Within the analog neural network block, we illustrate using a three-layer fully connected network, where the input and output dimensions are 2, and each hidden layer contains 14 nodes with bias, each layer features a resistive memory-based in-memory matrix multiplication module. This design utilizes shared negative weights to achieve a 50\% reduction of resistive memory cells. Additionally, \gls{tia}s are employed to translate the resultant current into voltages. Voltage activation is conducted by a \gls{relu} module, operational through a dual-diode setup within the \gls{tia} (with diode 1N4148). The processed outputs of analog neural network undergo further processes by a series of analog multipliers (AD633 from Analog Devices Inc., ADI), summing and inverting amplifiers, and integrators. The outputs are then fed back as inputs to the analog neural network block, thus creating a closed-loop system that is functionally equivalent to the neural differential equation within the diffusion framework. The system's inherent temporal progression mirrors the neural differential equation's continuous resolution. The system's versatility allows it to operate as a solver for either \gls{ode} or \gls{sde}, contingent upon whether noise is incorporated into the circuit. The external signals are generated by 12-bit DAC (MAX5742 from Analog Devices Inc./Maxim Integrated). Furthermore, the algorithm's requisite time and condition variables are incorporated at the \gls{tia}s' preceding current summation node, allowing for precise modulation and navigation of the diffusion trajectory. Details of the algorithm logic in PCB system can be found in Supplementary Fig.~4.

In the PCB system, the resistive memory array operates in two distinct modes: computation and programming. The switching between these modes is facilitated by single-pole double-throw (SPDT) switches (TMUX1134 from Texas Instruments, TI). When in computation mode, the resistive memory array is integrated into the analog score-based diffusion computational module. Conversely, in programming mode, the resistive memory array is interfaced with the programming circuit and connected to the B1500A Semiconductor Device Analyzer for operational manipulation. The detailed programming logic is delineated in Supplementary Fig.~3.

In the definition of score-based diffusion algorithm, the reverse generation process starts by sampling from a Gaussian distribution at time $T$ and ends at time $0$ to obtain the target distribution\cite{song2020score}. In the PCB experiment setup, the recorded $t = \SI{0}{\second}$ corresponds to the initial time $T$ in algorithm, and $t = \SI{1}{\second}$ corresponds to the end time $0$ in algorithm of the generation process.

\subsection*{In-memory score-based diffusion models}

In-memory score-based diffusion models are implementations of score-based diffusion models on resistive memory-based in-memory computing architectures, which have significant improvements in sampling speed and power consumption. Their software implementation primarily relies on the following methods.

\subsubsection*{Variance preserving score-based diffusion models}

The variance preserving score-based diffusion model's efficacy is showcased through its non-explosive variance across the diffusion timeline, with a linearly increasing $\beta (t)$ (from 0.001 to 0.5 for $t$ from 0 to $T$). The model's drift and diffusion coefficients are respectively encapsulated by\cite{song2020score}:

\begin{equation}
\boldsymbol{f}(\boldsymbol{x}, t) = -\frac{1}{2}\beta(t)\boldsymbol{x} \label{eq:drift},
\end{equation}

\begin{equation}
g(t) = \sqrt{\beta(t)} \label{eq:diffusion},
\end{equation}

This form of score-based diffusion model does not involve parameters with very large numerical values throughout the diffusion process, which facilitates convenient implementation on hardware.

\subsubsection*{Classifier-free guidance diffusion models}

Classifier-free guidance\cite{CFG} modifies the sampling process by adjusting the score function based on a trade-off between the data distribution and the conditional distribution of a specified condition. The method does not require a separate classifier model. The adjusted generation differential equation process can be expressed as:

\begin{equation}
    \frac{\mathrm{d}\boldsymbol{x}}{\mathrm{d} t} = \boldsymbol{f}(\boldsymbol{x}, t) - g^2(t) \boldsymbol{\Tilde{s}}_\theta(\boldsymbol{x}, \boldsymbol{c}, t) + g(t) \frac{\mathrm{d} \boldsymbol{w}}{\mathrm{d} t}
\label{conditional SDE}
\end{equation}

While the score function can be written by:

\begin{equation}
    \boldsymbol{\Tilde{s}}_\theta(\boldsymbol{x}, \boldsymbol{c}, t) = (1+\lambda)\boldsymbol{s}_\theta(\boldsymbol{x}, \boldsymbol{c}, t) - \lambda\boldsymbol{s}_\theta(\boldsymbol{x}, t)
\label{CFG}
\end{equation}

Where $\lambda$ is a hyperparameter that controls the strength of the guidance. Classifier-free guidance helps in steering the generative process toward samples that are more likely under certain conditions. The encoded signals of the condition term $c$ and time $t$ can be simultaneously added to the network on the hardware, providing guidance to the generation process.

\subsection*{KL divergence}

In the assessment of generative model quality, the \gls{kl} divergence metric is invoked to quantify the congruence between the distribution of the generated data and the distribution of the original data. The \gls{kl} divergence, expressed as follows:

\begin{equation}
D_{KL}(P | Q) = \sum_{x} P(x) \log \frac{P(x)}{Q(x)}\label{eq:KLdivergence}.
\end{equation}

Herein, $D_{KL}(P | Q)$ represents the information difference incurred when distribution $Q$ is employed to approximate the true distribution $P$. A small \gls{kl} divergence value is indicative of a close resemblance between $Q$ and $P$ in the probabilistic landscape. It quantifies the extent of deviation between the two distributions, with a divergence of 0 denoting perfect alignment. The \gls{kl} divergence between generated distribution and ground truth distribution measures the fidelity of generative models in replicating the statistical properties of the training dataset.

\subsection*{Time embedding module}

The time embedding module is an integral component for embedding time information into neural networks. It is mathematically formulated as:

\begin{equation}
\boldsymbol{v}_t = \text{TimeEmbedding}(t) = \left[\sin(2\pi\boldsymbol{W}t), \cos(2\pi\boldsymbol{W}t)\right],
\end{equation}

where $\boldsymbol{v}_t$ is the embedded vector, $t$ is the time input, and $\boldsymbol{W} \in \mathbb{R}^{d/2}$ is a vector of fixed parameters randomly initialized to scale the frequencies of the sine and cosine functions. This encoding process effectively captures the time signal in a format applied to the fully connected neural networks. By leveraging trigonometric projections, the module encodes time as a continuous and differentiable signal, which is significant for learning time dependencies within data. The resulting high-dimensional representation enriches the neural network with the ability to interpret and utilize time dynamics.

\subsection*{Training loss of \gls{vae}}

The loss used for training the \gls{vae} is defined as:

\begin{equation} 
    L_\text{VAE} = \text{MSE}(X, X') + \gamma \text{KL}(N(\mu_i, \sigma^2_i) , N(\hat{\mu_i}, 1)),  \quad i=0,1,\cdots, K,
\label{vae}
\end{equation}

where the first term, \(\mathrm{MSE}(X, X')\), represents the Mean Squared Error between the reconstructed outputs and the original inputs, serving as a metric for reconstruction fidelity. The second term, \(\mathrm{D_{KL}}(N(\mu_i, \sigma^2_i) \parallel N(\hat{\mu}_i, 1))\), measures the \gls{kl} divergence between the latent space's encoded distribution and the predefined target distribution, with the hyperparameter \(\gamma\) balancing the two loss components. Here, \(K\) signifies the number of label classes, and \(\hat{\mu}_i\) denotes the preset center for each class's distribution.

\subsection*{Data processing on \gls{emnist} dataset}

The Extended Modified National Institute of Standards and Technology (EMNIST) dataset is a comprehensive image classification repository comprising various classes representing English alphabet characters and numerals. Each image, encoded in grayscale, possesses a resolution of $28\times28$ pixels. The grayscale values are normalized to span a range of $[-1, 1]$. Subsequently, the images undergo downsampling, resulting in a reduced resolution of $14\times14$ pixels, and are further processed via center cropping to yield a final dimension of $12\times12$ pixels. Within this curated dataset, specific images that depict the letters \textit{H}, \textit{K}, and \textit{U} are segregated for focused analysis. The dataset of these images is used for the task of conditional latent score-based diffusion.

\section*{Acknowledgements}
This research is supported by the National Natural Science Foundation of China (Grant Nos. 62122004, 62374181), Hong Kong Research Grant Council (Grant Nos. 27206321, 17205922, 17212923). This research is also partially supported by ACCESS – AI Chip Center for Emerging Smart Systems, sponsored by Innovation and Technology Fund (ITF), Hong Kong SAR.


\section*{Competing interests}
The authors declare no competing interests.

\section*{Data availability}
The \gls{emnist} dataset is publicly available. All other measured data are freely available upon request. Source data are provided with this paper and also available at \url{https://github.com/yangjc97/memristor-sde-diffusion}.

\section*{Code availability}
The code that supports the plots within this paper and other findings of this study is available at \url{https://github.com/yangjc97/memristor-sde-diffusion}.
The code that supports the control of our in-memory neural differential equation solver printed circuit board system (see Supplementary Fig.~4 for control logic diagram) is available from the corresponding author on reasonable request.

\bibliography{ref}

\begin{thebibliography}{10}
\urlstyle{rm}
\expandafter\ifx\csname url\endcsname\relax
  \def\url#1{\texttt{#1}}\fi
\expandafter\ifx\csname urlprefix\endcsname\relax\def\urlprefix{URL }\fi
\expandafter\ifx\csname doiprefix\endcsname\relax\def\doiprefix{DOI: }\fi
\providecommand{\bibinfo}[2]{#2}
\providecommand{\eprint}[2][]{\url{#2}}

\bibitem{mullally2014memory}
\bibinfo{author}{Mullally, S.~L.} \& \bibinfo{author}{Maguire, E.~A.}
\newblock \bibinfo{journal}{\bibinfo{title}{Memory, imagination, and predicting the future: A common brain mechanism?}}
\newblock {\emph{\JournalTitle{The Neuroscientist}}} \textbf{\bibinfo{volume}{20}}, \bibinfo{pages}{220--234} (\bibinfo{year}{2014}).

\bibitem{videoworldsimulators2024}
\bibinfo{author}{Brooks, T.} \emph{et~al.}
\newblock \bibinfo{journal}{\bibinfo{title}{Video generation models as world simulators}}.
\newblock {\emph{\JournalTitle{OpenAI}}}  (\bibinfo{year}{2024}).

\bibitem{ho2020denoising}
\bibinfo{author}{Ho, J.}, \bibinfo{author}{Jain, A.} \& \bibinfo{author}{Abbeel, P.}
\newblock \bibinfo{journal}{\bibinfo{title}{Denoising diffusion probabilistic models}}.
\newblock {\emph{\JournalTitle{Advances in neural information processing systems}}} \textbf{\bibinfo{volume}{33}}, \bibinfo{pages}{6840--6851} (\bibinfo{year}{2020}).

\bibitem{karras}
\bibinfo{author}{Karras, T.}, \bibinfo{author}{Aittala, M.}, \bibinfo{author}{Aila, T.} \& \bibinfo{author}{Laine, S.}
\newblock \bibinfo{journal}{\bibinfo{title}{Elucidating the design space of diffusion-based generative models}}.
\newblock {\emph{\JournalTitle{arXiv preprint arXiv:2206.00364}}}  (\bibinfo{year}{2022}).

\bibitem{SD}
\bibinfo{author}{Rombach, R.}, \bibinfo{author}{Blattmann, A.}, \bibinfo{author}{Lorenz, D.}, \bibinfo{author}{Esser, P.} \& \bibinfo{author}{Ommer, B.}
\newblock \bibinfo{title}{High-resolution image synthesis with latent diffusion models}.
\newblock In \emph{\bibinfo{booktitle}{Proceedings of the IEEE/CVF Conference on Computer Vision and Pattern Recognition}}, \bibinfo{pages}{10684--10695} (\bibinfo{year}{2022}).

\bibitem{ViD4k}
\bibinfo{author}{Chen, J.} \emph{et~al.}
\newblock \bibinfo{journal}{\bibinfo{title}{Pixart-sigma: Weak-to-strong training of diffusion transformer for 4k text-to-image generation}}.
\newblock {\emph{\JournalTitle{arXiv preprint arXiv:2403.04692}}}  (\bibinfo{year}{2024}).

\bibitem{a1}
\bibinfo{author}{Kazerouni, A.} \emph{et~al.}
\newblock \bibinfo{journal}{\bibinfo{title}{Diffusion models for medical image analysis: A comprehensive survey}}.
\newblock {\emph{\JournalTitle{arXiv preprint arXiv:2211.07804}}}  (\bibinfo{year}{2022}).

\bibitem{a3}
\bibinfo{author}{Yang, L.}, \bibinfo{author}{Zhang, Z.} \& \bibinfo{author}{Hong, S.}
\newblock \bibinfo{journal}{\bibinfo{title}{Diffusion models: A comprehensive survey of methods and applications}}.
\newblock {\emph{\JournalTitle{arXiv preprint arXiv:2209.00796}}}  (\bibinfo{year}{2022}).

\bibitem{a9}
\bibinfo{author}{Fan, W.} \emph{et~al.}
\newblock \bibinfo{journal}{\bibinfo{title}{Generative diffusion models on graphs: Methods and applications}}.
\newblock {\emph{\JournalTitle{arXiv preprint arXiv:2302.02591}}}  (\bibinfo{year}{2023}).

\bibitem{RN94}
\bibinfo{author}{Zhang, L.} \& \bibinfo{author}{Agrawala, M.}
\newblock \bibinfo{journal}{\bibinfo{title}{Adding conditional control to text-to-image diffusion models}}.
\newblock {\emph{\JournalTitle{arXiv preprint arXiv:2302.05543}}}  (\bibinfo{year}{2023}).

\bibitem{a2}
\bibinfo{author}{Zhu, Y.} \& \bibinfo{author}{Zhao, Y.}
\newblock \bibinfo{journal}{\bibinfo{title}{Diffusion models in nlp: A survey}}.
\newblock {\emph{\JournalTitle{arXiv preprint arXiv:2303.07576}}}  (\bibinfo{year}{2023}).

\bibitem{a6_3d}
\bibinfo{author}{Poole, B.}, \bibinfo{author}{Jain, A.}, \bibinfo{author}{Barron, J.~T.} \& \bibinfo{author}{Mildenhall, B.}
\newblock \bibinfo{journal}{\bibinfo{title}{Dreamfusion: Text-to-3d using 2d diffusion}}.
\newblock {\emph{\JournalTitle{arXiv preprint arXiv:2209.14988}}}  (\bibinfo{year}{2022}).

\bibitem{song2020score}
\bibinfo{author}{Song, Y.} \emph{et~al.}
\newblock \bibinfo{journal}{\bibinfo{title}{Score-based generative modeling through stochastic differential equations}}.
\newblock {\emph{\JournalTitle{arXiv preprint arXiv:2011.13456}}}  (\bibinfo{year}{2020}).

\bibitem{b1}
\bibinfo{author}{Song, Y.}, \bibinfo{author}{Dhariwal, P.}, \bibinfo{author}{Chen, M.} \& \bibinfo{author}{Sutskever, I.}
\newblock \bibinfo{journal}{\bibinfo{title}{Consistency models}}.
\newblock {\emph{\JournalTitle{arXiv preprint arXiv:2303.01469}}}  (\bibinfo{year}{2023}).

\bibitem{b2}
\bibinfo{author}{Song, Y.} \& \bibinfo{author}{Ermon, S.}
\newblock \bibinfo{journal}{\bibinfo{title}{Generative modeling by estimating gradients of the data distribution}}.
\newblock {\emph{\JournalTitle{Advances in Neural Information Processing Systems}}} \textbf{\bibinfo{volume}{32}} (\bibinfo{year}{2019}).

\bibitem{b4}
\bibinfo{author}{Song, Y.} \& \bibinfo{author}{Ermon, S.}
\newblock \bibinfo{journal}{\bibinfo{title}{Improved techniques for training score-based generative models}}.
\newblock {\emph{\JournalTitle{Advances in neural information processing systems}}} \textbf{\bibinfo{volume}{33}}, \bibinfo{pages}{12438--12448} (\bibinfo{year}{2020}).

\bibitem{b3}
\bibinfo{author}{Ghimire, S.} \emph{et~al.}
\newblock \bibinfo{journal}{\bibinfo{title}{Geometry of score based generative models}}.
\newblock {\emph{\JournalTitle{arXiv preprint arXiv:2302.04411}}}  (\bibinfo{year}{2023}).

\bibitem{b6}
\bibinfo{author}{Meng, C.} \emph{et~al.}
\newblock \bibinfo{journal}{\bibinfo{title}{Sdedit: Guided image synthesis and editing with stochastic differential equations}}.
\newblock {\emph{\JournalTitle{arXiv preprint arXiv:2108.01073}}}  (\bibinfo{year}{2021}).

\bibitem{b11}
\bibinfo{author}{Chen, R.~T.}, \bibinfo{author}{Rubanova, Y.}, \bibinfo{author}{Bettencourt, J.} \& \bibinfo{author}{Duvenaud, D.~K.}
\newblock \bibinfo{journal}{\bibinfo{title}{Neural ordinary differential equations}}.
\newblock {\emph{\JournalTitle{Advances in neural information processing systems}}} \textbf{\bibinfo{volume}{31}} (\bibinfo{year}{2018}).

\bibitem{b12}
\bibinfo{author}{Liu, X.} \emph{et~al.}
\newblock \bibinfo{journal}{\bibinfo{title}{Neural sde: Stabilizing neural ode networks with stochastic noise}}.
\newblock {\emph{\JournalTitle{arXiv preprint arXiv:1906.02355}}}  (\bibinfo{year}{2019}).

\bibitem{b13}
\bibinfo{author}{Tzen, B.} \& \bibinfo{author}{Raginsky, M.}
\newblock \bibinfo{journal}{\bibinfo{title}{Neural stochastic differential equations: Deep latent gaussian models in the diffusion limit}}.
\newblock {\emph{\JournalTitle{arXiv preprint arXiv:1905.09883}}}  (\bibinfo{year}{2019}).

\bibitem{rc2}
\bibinfo{author}{Theis, T.~N.} \& \bibinfo{author}{Wong, H.-S.~P.}
\newblock \bibinfo{journal}{\bibinfo{title}{The end of moore's law: A new beginning for information technology}}.
\newblock {\emph{\JournalTitle{Computing in science \& engineering}}} \textbf{\bibinfo{volume}{19}}, \bibinfo{pages}{41--50} (\bibinfo{year}{2017}).

\bibitem{s1}
\bibinfo{author}{Ye, W.} \emph{et~al.}
\newblock \bibinfo{journal}{\bibinfo{title}{A 28-nm rram computing-in-memory macro using weighted hybrid 2t1r cell array and reference subtracting sense amplifier for ai edge inference}}.
\newblock {\emph{\JournalTitle{IEEE Journal of Solid-State Circuits}}}  (\bibinfo{year}{2023}).

\bibitem{s5}
\bibinfo{author}{Li, C.} \emph{et~al.}
\newblock \bibinfo{journal}{\bibinfo{title}{Analogue signal and image processing with large memristor crossbars}}.
\newblock {\emph{\JournalTitle{Nature electronics}}} \textbf{\bibinfo{volume}{1}}, \bibinfo{pages}{52--59} (\bibinfo{year}{2018}).

\bibitem{s9}
\bibinfo{author}{Zhou, P.}, \bibinfo{author}{Choi, D.-U.}, \bibinfo{author}{Lu, W.~D.}, \bibinfo{author}{Kang, S.-M.} \& \bibinfo{author}{Eshraghian, J.~K.}
\newblock \bibinfo{journal}{\bibinfo{title}{Gradient-based neuromorphic learning on dynamical rram arrays}}.
\newblock {\emph{\JournalTitle{IEEE Journal on Emerging and Selected Topics in Circuits and Systems}}} \textbf{\bibinfo{volume}{12}}, \bibinfo{pages}{888--897} (\bibinfo{year}{2022}).

\bibitem{s10}
\bibinfo{author}{Wang, R.} \emph{et~al.}
\newblock \bibinfo{journal}{\bibinfo{title}{Implementing in-situ self-organizing maps with memristor crossbar arrays for data mining and optimization}}.
\newblock {\emph{\JournalTitle{Nature communications}}} \textbf{\bibinfo{volume}{13}}, \bibinfo{pages}{2289} (\bibinfo{year}{2022}).

\bibitem{sun2023full}
\bibinfo{author}{Sun, Z.} \emph{et~al.}
\newblock \bibinfo{journal}{\bibinfo{title}{A full spectrum of computing-in-memory technologies}}.
\newblock {\emph{\JournalTitle{Nature Electronics}}} \textbf{\bibinfo{volume}{6}}, \bibinfo{pages}{823--835} (\bibinfo{year}{2023}).

\bibitem{acc2}
\bibinfo{author}{Lu, C.} \emph{et~al.}
\newblock \bibinfo{journal}{\bibinfo{title}{Dpm-solver++: Fast solver for guided sampling of diffusion probabilistic models}}.
\newblock {\emph{\JournalTitle{arXiv preprint arXiv:2211.01095}}}  (\bibinfo{year}{2022}).

\bibitem{acc3}
\bibinfo{author}{Zhou, Z.}, \bibinfo{author}{Chen, D.}, \bibinfo{author}{Wang, C.} \& \bibinfo{author}{Chen, C.}
\newblock \bibinfo{journal}{\bibinfo{title}{Fast ode-based sampling for diffusion models in around 5 steps}}.
\newblock {\emph{\JournalTitle{arXiv preprint arXiv:2312.00094}}}  (\bibinfo{year}{2023}).

\bibitem{acc4}
\bibinfo{author}{Lipman, Y.}, \bibinfo{author}{Chen, R.~T.}, \bibinfo{author}{Ben-Hamu, H.}, \bibinfo{author}{Nickel, M.} \& \bibinfo{author}{Le, M.}
\newblock \bibinfo{journal}{\bibinfo{title}{Flow matching for generative modeling}}.
\newblock {\emph{\JournalTitle{arXiv preprint arXiv:2210.02747}}}  (\bibinfo{year}{2022}).

\bibitem{acc5}
\bibinfo{author}{Liu, X.}, \bibinfo{author}{Gong, C.} \& \bibinfo{author}{Liu, Q.}
\newblock \bibinfo{journal}{\bibinfo{title}{Flow straight and fast: Learning to generate and transfer data with rectified flow}}.
\newblock {\emph{\JournalTitle{arXiv preprint arXiv:2209.03003}}}  (\bibinfo{year}{2022}).

\bibitem{acc6}
\bibinfo{author}{Nichol, A.~Q.} \& \bibinfo{author}{Dhariwal, P.}
\newblock \bibinfo{title}{Improved denoising diffusion probabilistic models}.
\newblock In \emph{\bibinfo{booktitle}{International Conference on Machine Learning}}, \bibinfo{pages}{8162--8171} (\bibinfo{publisher}{PMLR}, \bibinfo{year}{2021}).

\bibitem{acc10}
\bibinfo{author}{Liu, L.}, \bibinfo{author}{Ren, Y.}, \bibinfo{author}{Lin, Z.} \& \bibinfo{author}{Zhao, Z.}
\newblock \bibinfo{journal}{\bibinfo{title}{Pseudo numerical methods for diffusion models on manifolds}}.
\newblock {\emph{\JournalTitle{arXiv preprint arXiv:2202.09778}}}  (\bibinfo{year}{2022}).

\bibitem{acc11}
\bibinfo{author}{Kong, L.}, \bibinfo{author}{Sun, J.} \& \bibinfo{author}{Zhang, C.}
\newblock \bibinfo{journal}{\bibinfo{title}{Sde-net: Equipping deep neural networks with uncertainty estimates}}.
\newblock {\emph{\JournalTitle{arXiv preprint arXiv:2008.10546}}}  (\bibinfo{year}{2020}).

\bibitem{rc1}
\bibinfo{author}{Wan, W.} \emph{et~al.}
\newblock \bibinfo{journal}{\bibinfo{title}{A compute-in-memory chip based on resistive random-access memory}}.
\newblock {\emph{\JournalTitle{Nature}}} \textbf{\bibinfo{volume}{608}}, \bibinfo{pages}{504--512}, \doiprefix\url{10.1038/s41586-022-04992-8} (\bibinfo{year}{2022}).

\bibitem{rc4}
\bibinfo{author}{Wang, Z.} \emph{et~al.}
\newblock \bibinfo{journal}{\bibinfo{title}{Fully memristive neural networks for pattern classification with unsupervised learning}}.
\newblock {\emph{\JournalTitle{Nature Electronics}}} \textbf{\bibinfo{volume}{1}}, \bibinfo{pages}{137--145} (\bibinfo{year}{2018}).

\bibitem{rc5}
\bibinfo{author}{James, A.~P.}
\newblock \bibinfo{journal}{\bibinfo{title}{A hybrid memristor-cmos chip for ai}}.
\newblock {\emph{\JournalTitle{Nature Electronics}}} \textbf{\bibinfo{volume}{2}}, \bibinfo{pages}{268--269}, \doiprefix\url{10.1038/s41928-019-0274-6} (\bibinfo{year}{2019}).
\newblock \bibinfo{note}{Ik8ab Times Cited:11 Cited References Count:7}.

\bibitem{rc6}
\bibinfo{author}{Ielmini, D.} \& \bibinfo{author}{Wong, H.-S.~P.}
\newblock \bibinfo{journal}{\bibinfo{title}{In-memory computing with resistive switching devices}}.
\newblock {\emph{\JournalTitle{Nature electronics}}} \textbf{\bibinfo{volume}{1}}, \bibinfo{pages}{333--343} (\bibinfo{year}{2018}).

\bibitem{rc9}
\bibinfo{author}{Strukov, D.~B.}, \bibinfo{author}{Snider, G.~S.}, \bibinfo{author}{Stewart, D.~R.} \& \bibinfo{author}{Williams, R.~S.}
\newblock \bibinfo{journal}{\bibinfo{title}{The missing memristor found}}.
\newblock {\emph{\JournalTitle{Nature}}} \textbf{\bibinfo{volume}{453}}, \bibinfo{pages}{80--83}, \doiprefix\url{10.1038/nature06932} (\bibinfo{year}{2008}).

\bibitem{yan2023memristive}
\bibinfo{author}{Yan, B.}, \bibinfo{author}{Yang, Y.} \& \bibinfo{author}{Huang, R.}
\newblock \bibinfo{journal}{\bibinfo{title}{Memristive dynamics enabled neuromorphic computing systems}}.
\newblock {\emph{\JournalTitle{Science China Information Sciences}}} \textbf{\bibinfo{volume}{66}}, \bibinfo{pages}{200401} (\bibinfo{year}{2023}).

\bibitem{rc7}
\bibinfo{author}{Li, C.} \emph{et~al.}
\newblock \bibinfo{journal}{\bibinfo{title}{Long short-term memory networks in memristor crossbar arrays}}.
\newblock {\emph{\JournalTitle{Nature Machine Intelligence}}} \textbf{\bibinfo{volume}{1}}, \bibinfo{pages}{49--57} (\bibinfo{year}{2019}).

\bibitem{rc8}
\bibinfo{author}{Lanza, M.} \emph{et~al.}
\newblock \bibinfo{journal}{\bibinfo{title}{Memristive technologies for data storage, computation, encryption, and radio-frequency communication}}.
\newblock {\emph{\JournalTitle{Science}}} \textbf{\bibinfo{volume}{376}}, \bibinfo{pages}{eabj9979}, \doiprefix\url{doi:10.1126/science.abj9979} (\bibinfo{year}{2022}).

\bibitem{SVM}
\bibinfo{author}{Yan, X.} \emph{et~al.}
\newblock \bibinfo{journal}{\bibinfo{title}{Reconfigurable mixed-kernel heterojunction transistors for personalized support vector machine classification}}.
\newblock {\emph{\JournalTitle{Nature Electronics}}} \textbf{\bibinfo{volume}{6}}, \bibinfo{pages}{862--869} (\bibinfo{year}{2023}).

\bibitem{rc10}
\bibinfo{author}{Wang, Z.} \emph{et~al.}
\newblock \bibinfo{journal}{\bibinfo{title}{Resistive switching materials for information processing}}.
\newblock {\emph{\JournalTitle{Nature Reviews Materials}}} \textbf{\bibinfo{volume}{5}}, \bibinfo{pages}{173--195} (\bibinfo{year}{2020}).

\bibitem{s6}
\bibinfo{author}{Zhang, X.} \emph{et~al.}
\newblock \bibinfo{journal}{\bibinfo{title}{An artificial spiking afferent nerve based on mott memristors for neurorobotics}}.
\newblock {\emph{\JournalTitle{Nature communications}}} \textbf{\bibinfo{volume}{11}}, \bibinfo{pages}{51} (\bibinfo{year}{2020}).

\bibitem{s7}
\bibinfo{author}{Kumar, S.}, \bibinfo{author}{Wang, X.}, \bibinfo{author}{Strachan, J.~P.}, \bibinfo{author}{Yang, Y.} \& \bibinfo{author}{Lu, W.~D.}
\newblock \bibinfo{journal}{\bibinfo{title}{Dynamical memristors for higher-complexity neuromorphic computing}}.
\newblock {\emph{\JournalTitle{Nature Reviews Materials}}} \textbf{\bibinfo{volume}{7}}, \bibinfo{pages}{575--591} (\bibinfo{year}{2022}).

\bibitem{s8}
\bibinfo{author}{Chen, S.}, \bibinfo{author}{Zhang, T.}, \bibinfo{author}{Tappertzhofen, S.}, \bibinfo{author}{Yang, Y.} \& \bibinfo{author}{Valov, I.}
\newblock \bibinfo{journal}{\bibinfo{title}{Electrochemical‐memristor‐based artificial neurons and synapses—fundamentals, applications, and challenges}}.
\newblock {\emph{\JournalTitle{Advanced Materials}}} \textbf{\bibinfo{volume}{35}}, \bibinfo{pages}{2301924} (\bibinfo{year}{2023}).

\bibitem{s11}
\bibinfo{author}{Sebastian, A.}, \bibinfo{author}{Le~Gallo, M.}, \bibinfo{author}{Khaddam-Aljameh, R.} \& \bibinfo{author}{Eleftheriou, E.}
\newblock \bibinfo{journal}{\bibinfo{title}{Memory devices and applications for in-memory computing}}.
\newblock {\emph{\JournalTitle{Nature nanotechnology}}} \textbf{\bibinfo{volume}{15}}, \bibinfo{pages}{529--544} (\bibinfo{year}{2020}).

\bibitem{s12}
\bibinfo{author}{Xia, Q.} \& \bibinfo{author}{Yang, J.~J.}
\newblock \bibinfo{journal}{\bibinfo{title}{Memristive crossbar arrays for brain-inspired computing}}.
\newblock {\emph{\JournalTitle{Nature materials}}} \textbf{\bibinfo{volume}{18}}, \bibinfo{pages}{309--323} (\bibinfo{year}{2019}).

\bibitem{moon2019temporal}
\bibinfo{author}{Moon, J.} \emph{et~al.}
\newblock \bibinfo{journal}{\bibinfo{title}{Temporal data classification and forecasting using a memristor-based reservoir computing system}}.
\newblock {\emph{\JournalTitle{Nature Electronics}}} \textbf{\bibinfo{volume}{2}}, \bibinfo{pages}{480--487} (\bibinfo{year}{2019}).

\bibitem{yang2020retransformer}
\bibinfo{author}{Yang, X.}, \bibinfo{author}{Yan, B.}, \bibinfo{author}{Li, H.} \& \bibinfo{author}{Chen, Y.}
\newblock \bibinfo{title}{Retransformer: Reram-based processing-in-memory architecture for transformer acceleration}.
\newblock In \emph{\bibinfo{booktitle}{Proceedings of the 39th International Conference on Computer-Aided Design}}, \bibinfo{pages}{1--9} (\bibinfo{year}{2020}).

\bibitem{zhu2020comprehensive}
\bibinfo{author}{Zhu, J.}, \bibinfo{author}{Zhang, T.}, \bibinfo{author}{Yang, Y.} \& \bibinfo{author}{Huang, R.}
\newblock \bibinfo{journal}{\bibinfo{title}{A comprehensive review on emerging artificial neuromorphic devices}}.
\newblock {\emph{\JournalTitle{Applied Physics Reviews}}} \textbf{\bibinfo{volume}{7}} (\bibinfo{year}{2020}).

\bibitem{CFG}
\bibinfo{author}{Ho, J.} \& \bibinfo{author}{Salimans, T.}
\newblock \bibinfo{journal}{\bibinfo{title}{Classifier-free diffusion guidance}}.
\newblock {\emph{\JournalTitle{arXiv preprint arXiv:2207.12598}}}  (\bibinfo{year}{2022}).

\bibitem{noise2}
\bibinfo{author}{Wang, D.} \emph{et~al.}
\newblock \bibinfo{title}{{Stochastic Emerging Resistive Memories for Unconventional Computing}}.
\newblock In \emph{\bibinfo{booktitle}{{Advanced Memory Technology: Functional Materials and Devices}}}, \doiprefix\url{10.1039/BK9781839169946-00240} (\bibinfo{publisher}{Royal Society of Chemistry}, \bibinfo{year}{2023}).

\bibitem{yang2022tolerating}
\bibinfo{author}{Yang, X.}, \bibinfo{author}{Wu, C.}, \bibinfo{author}{Li, M.} \& \bibinfo{author}{Chen, Y.}
\newblock \bibinfo{journal}{\bibinfo{title}{Tolerating noise effects in processing-in-memory systems for neural networks: A hardware--software codesign perspective}}.
\newblock {\emph{\JournalTitle{Advanced Intelligent Systems}}} \textbf{\bibinfo{volume}{4}}, \bibinfo{pages}{2200029} (\bibinfo{year}{2022}).

\bibitem{wei2023three}
\bibinfo{author}{Wei, T.} \emph{et~al.}
\newblock \bibinfo{journal}{\bibinfo{title}{Three-dimensional reconstruction of conductive filaments in hfox-based memristor}}.
\newblock {\emph{\JournalTitle{Advanced Materials}}} \textbf{\bibinfo{volume}{35}}, \bibinfo{pages}{2209925} (\bibinfo{year}{2023}).

\bibitem{cai2020power}
\bibinfo{author}{Cai, F.} \emph{et~al.}
\newblock \bibinfo{journal}{\bibinfo{title}{Power-efficient combinatorial optimization using intrinsic noise in memristor hopfield neural networks}}.
\newblock {\emph{\JournalTitle{Nature Electronics}}} \textbf{\bibinfo{volume}{3}}, \bibinfo{pages}{409--418} (\bibinfo{year}{2020}).

\bibitem{modell2003imagination}
\bibinfo{author}{Modell, A.~H.}
\newblock \emph{\bibinfo{title}{Imagination and the meaningful brain}} (\bibinfo{publisher}{mit Press}, \bibinfo{year}{2003}).

\bibitem{rr2}
\bibinfo{author}{Rao, M.} \emph{et~al.}
\newblock \bibinfo{journal}{\bibinfo{title}{Thousands of conductance levels in memristors integrated on cmos}}.
\newblock {\emph{\JournalTitle{Nature}}} \textbf{\bibinfo{volume}{615}}, \bibinfo{pages}{823--829} (\bibinfo{year}{2023}).

\bibitem{s2}
\bibinfo{author}{Jiang, H.}, \bibinfo{author}{Li, W.}, \bibinfo{author}{Huang, S.} \& \bibinfo{author}{Yu, S.}
\newblock \bibinfo{title}{A 40nm analog-input adc-free compute-in-memory rram macro with pulse-width modulation between sub-arrays}.
\newblock In \emph{\bibinfo{booktitle}{2022 IEEE Symposium on VLSI Technology and Circuits (VLSI Technology and Circuits)}}, \bibinfo{pages}{266--267} (\bibinfo{publisher}{IEEE}, \bibinfo{year}{2022}).

\bibitem{s3}
\bibinfo{author}{Le~Gallo, M.} \emph{et~al.}
\newblock \bibinfo{journal}{\bibinfo{title}{A 64-core mixed-signal in-memory compute chip based on phase-change memory for deep neural network inference}}.
\newblock {\emph{\JournalTitle{Nature Electronics}}} \textbf{\bibinfo{volume}{6}}, \bibinfo{pages}{680--693} (\bibinfo{year}{2023}).

\bibitem{s4}
\bibinfo{author}{Li, Y.} \emph{et~al.}
\newblock \bibinfo{journal}{\bibinfo{title}{An adc-less rram-based computing-in-memory macro with binary cnn for efficient edge ai}}.
\newblock {\emph{\JournalTitle{IEEE Transactions on Circuits and Systems II: Express Briefs}}}  (\bibinfo{year}{2023}).

\bibitem{wang2023analog}
\bibinfo{author}{Wang, S.} \emph{et~al.}
\newblock \bibinfo{journal}{\bibinfo{title}{In-memory analog solution of compressed sensing recovery in one step}}.
\newblock {\emph{\JournalTitle{Science Advances}}} \textbf{\bibinfo{volume}{9}}, \bibinfo{pages}{eadj2908}, \doiprefix\url{10.1126/sciadv.adj2908} (\bibinfo{year}{2023}).

\bibitem{ielmini2021brain}
\bibinfo{author}{Ielmini, D.}, \bibinfo{author}{Wang, Z.} \& \bibinfo{author}{Liu, Y.}
\newblock \bibinfo{journal}{\bibinfo{title}{Brain-inspired computing via memory device physics}}.
\newblock {\emph{\JournalTitle{APL Materials}}} \textbf{\bibinfo{volume}{9}} (\bibinfo{year}{2021}).

\bibitem{rr1}
\bibinfo{author}{Song, M.-K.} \emph{et~al.}
\newblock \bibinfo{journal}{\bibinfo{title}{Recent advances and future prospects for memristive materials, devices, and systems}}.
\newblock {\emph{\JournalTitle{ACS nano}}} \textbf{\bibinfo{volume}{17}}, \bibinfo{pages}{11994--12039} (\bibinfo{year}{2023}).

\bibitem{c12}
\bibinfo{author}{Wang, Z.} \emph{et~al.}
\newblock \bibinfo{journal}{\bibinfo{title}{Reinforcement learning with analogue memristor arrays}}.
\newblock {\emph{\JournalTitle{Nature electronics}}} \textbf{\bibinfo{volume}{2}}, \bibinfo{pages}{115--124} (\bibinfo{year}{2019}).

\bibitem{c1}
\bibinfo{author}{Wang, S.} \emph{et~al.}
\newblock \bibinfo{journal}{\bibinfo{title}{Convolutional echo‐state network with random memristors for spatiotemporal signal classification}}.
\newblock {\emph{\JournalTitle{Advanced Intelligent Systems}}} \bibinfo{pages}{2200027} (\bibinfo{year}{2022}).

\bibitem{yi2023activity}
\bibinfo{author}{Yi, S.-i.}, \bibinfo{author}{Kendall, J.~D.}, \bibinfo{author}{Williams, R.~S.} \& \bibinfo{author}{Kumar, S.}
\newblock \bibinfo{journal}{\bibinfo{title}{Activity-difference training of deep neural networks using memristor crossbars}}.
\newblock {\emph{\JournalTitle{Nature Electronics}}} \textbf{\bibinfo{volume}{6}}, \bibinfo{pages}{45--51} (\bibinfo{year}{2023}).

\bibitem{sun2020one}
\bibinfo{author}{Sun, Z.}, \bibinfo{author}{Pedretti, G.}, \bibinfo{author}{Bricalli, A.} \& \bibinfo{author}{Ielmini, D.}
\newblock \bibinfo{journal}{\bibinfo{title}{One-step regression and classification with cross-point resistive memory arrays}}.
\newblock {\emph{\JournalTitle{Science advances}}} \textbf{\bibinfo{volume}{6}}, \bibinfo{pages}{eaay2378} (\bibinfo{year}{2020}).

\bibitem{zhang2023edge}
\bibinfo{author}{Zhang, W.} \emph{et~al.}
\newblock \bibinfo{journal}{\bibinfo{title}{Edge learning using a fully integrated neuro-inspired memristor chip}}.
\newblock {\emph{\JournalTitle{Science}}} \textbf{\bibinfo{volume}{381}}, \bibinfo{pages}{1205--1211} (\bibinfo{year}{2023}).

\bibitem{milo2021accurate}
\bibinfo{author}{Milo, V.} \emph{et~al.}
\newblock \bibinfo{journal}{\bibinfo{title}{Accurate program/verify schemes of resistive switching memory (rram) for in-memory neural network circuits}}.
\newblock {\emph{\JournalTitle{IEEE Transactions on Electron Devices}}} \textbf{\bibinfo{volume}{68}}, \bibinfo{pages}{3832--3837} (\bibinfo{year}{2021}).

\bibitem{fully}
\bibinfo{author}{Kiani, F.}, \bibinfo{author}{Yin, J.}, \bibinfo{author}{Wang, Z.}, \bibinfo{author}{Yang, J.~J.} \& \bibinfo{author}{Xia, Q.}
\newblock \bibinfo{journal}{\bibinfo{title}{A fully hardware-based memristive multilayer neural network}}.
\newblock {\emph{\JournalTitle{Science advances}}} \textbf{\bibinfo{volume}{7}}, \bibinfo{pages}{eabj4801} (\bibinfo{year}{2021}).

\bibitem{comp}
\bibinfo{author}{Xie, S.} \emph{et~al.}
\newblock \bibinfo{title}{16.2 edram-cim: Compute-in-memory design with reconfigurable embedded-dynamic-memory array realizing adaptive data converters and charge-domain computing}.
\newblock In \emph{\bibinfo{booktitle}{2021 IEEE International Solid-State Circuits Conference (ISSCC)}}, vol.~\bibinfo{volume}{64}, \bibinfo{pages}{248--250} (\bibinfo{publisher}{IEEE}, \bibinfo{year}{2021}).

\bibitem{VAE}
\bibinfo{author}{Kingma, D.~P.} \& \bibinfo{author}{Welling, M.}
\newblock \bibinfo{journal}{\bibinfo{title}{Auto-encoding variational bayes}}.
\newblock {\emph{\JournalTitle{arXiv preprint arXiv:1312.6114}}}  (\bibinfo{year}{2013}).

\bibitem{noise1}
\bibinfo{author}{Kandiah, K.} \& \bibinfo{author}{Whiting, F.}
\newblock \bibinfo{journal}{\bibinfo{title}{Low frequency noise in junction field effect transistors}}.
\newblock {\emph{\JournalTitle{Solid-State Electronics}}} \textbf{\bibinfo{volume}{21}}, \bibinfo{pages}{1079--1088} (\bibinfo{year}{1978}).

\end{thebibliography}

\end{document}